\documentclass{llncs}
\usepackage{epsfig,epic,eepic,amssymb,latexsym}
\usepackage{gastex}

\renewcommand{\part}{\sim}

\newcommand{\motnouv}[1]{{\em #1}}
\newcommand{\R}{\mathbb{R}}
\newcommand{\Q}{\mathbb{Q}}

\newcommand{\ELAN}{{\sf ELAN}}
\newcounter{largeurul}
\newcounter{hauteurul}
\newcounter{marginul}
\newcounter{l}
\newcounter{h}
\newcounter{ox}
\newcounter{oy}
\gasset{Nadjust=w,Nadjustdist=2}
\newcommand{\initialnode}{\node[Nmarks=r]}

\newcommand{\edgeleft}{\drawedge[ELside=l]}
\newcommand{\edgeright}{\drawedge[ELside=r]}
\newcommand{\edgecurveleft}{\drawedge[curvedepth=3]}

\newcommand{\edgecond}[3]{ \edgeleft(#1){#2} \edgeright(#1){#3} }
\newcommand{\edgecondcurve}[3] {\edgecurveleft(#1){#2 #3}}

\newcommand{\conc}[2]{\begin{tabular}{c} #1 \\ #2 \end{tabular}}
\newenvironment{automaton}[4]{
  \setcounter{largeurul}{#1}
  \setcounter{hauteurul}{#2}
  \setcounter{l}{\thelargeurul+2*\themarginul}
  \setcounter{h}{\thehauteurul+2*\themarginul}
  \setcounter{ox}{-\themarginul}
  \setcounter{oy}{-\themarginul}
  \setlength{\unitlength}{#4/\value{#3}}
  \begin{center}
    \begin{picture}(\thel,\theh)(\theox,\theoy)
}{
    \end{picture}
  \end{center}
}

\newcommand{\lsemantics}{[[}
\newcommand{\rsemantics}{]]}

\begin{document}

\newcommand{\LORIAaddress}{{\sc LORIA \& INRIA}, 615 rue du Jardin Botanique,
                BP 101, 54602 Villers-l{\`e}s-Nancy Cedex, Nancy, France. {\bf
  \{Olivier.Bournez,Hassen.Kacem,Claude.Kirchner\}@loria.fr}}

\newcommand{\ENSaddress}{{\sc ENS-Lyon}, 46 All{\'e}e d'Italie, 69364 Lyon
  Cedex 07, France. {\bf Emmanuel.Beffara@ens-lyon.fr}}

\title{Verification of Timed Automata Using Rewrite Rules and Strategies}
\author{Emmanuel Beffara\inst{2} \and Olivier Bournez\inst{1} \and Hassen Kacem\inst{1}
  \and Claude Kirchner\inst{1}}
\institute{\LORIAaddress \and \ENSaddress}
\date{}

\maketitle

\begin{abstract}
  \ELAN\ is a powerful language and environment for specifying and
  prototyping deduction systems in a language based on rewrite rules
  controlled by strategies.  Timed automata is a class of continuous
  real-time models of reactive systems for which efficient model-checking
  algorithms have been devised.  In this paper, we show that these
  algorithms can very easily be prototyped in the \ELAN\ system.
  
  This paper argues through this example that rewriting based systems
  relying on rules \textit{and} strategies are a good framework to
  prototype, study and test rather efficiently symbolic model-checking
  algorithms, i.e.  algorithms which involve combination of graph
  exploration rules, deduction rules, constraint solving techniques
  and decision procedures.
\end{abstract}

  

\section{Introduction}

\ELAN\ is a powerful language and environment for specifying and
prototyping deduction systems in a language based on rewrite rules
controlled by strategies. It offers a natural and simple framework for
the combination of the computation and the deduction paradigms. The
logical and semantical foundations of the \ELAN\ system rely
respectively on rewriting logic~\cite{MeseguerTCS92} and rewriting
calculus~\cite{CirsteaKirchner-LivreFroCoS99} and are in particular
described in \cite{ELAN-wrla98,CirsteaThese2000}.

Timed automata \cite{AD94} is a particular class of hybrid systems,
i.e.\ systems consisting of a mixture of continuous evolutions and
discrete transitions. They can be seen as automata augmented with
clock variables, which can be reset to $0$ by guarded transitions of
some special type. They have proven to be a very useful formalism for
describing timed systems, for which verification and synthesis
algorithms exist \cite{Alu98,AD94}, and are implemented in several
model-checking tools such as {\sf TIMED-COSPAN} \cite{AK95}, {\sf KRONOS}
\cite{DOTY95} or {\sf UPPAAL} \cite{LPY97}.

This paper describes our  experience using the \ELAN\
system to prototype the reachability verification algorithms implemented in the
model-checking tools for timed automata.

It is known that rewriting logic is a good framework for unifying the
different models of discrete-time reactive systems
\cite{MeseguerTCS92}.  Rewriting logic can be extended to deal with
continuous real-time models. Such an extension, called ``Timed
rewriting logic'' has been investigated, and applied to several
examples and specification languages \cite{KW97,PKW96,SK00}.  In this
approach the time is somehow built in the logic. Another approach is
to express continuous real-time models directly in rewriting logic.
This has been investigated in \cite{OM96,OM99} and recently Olveczky
and Meseguer have conceived ``Real-Time Maude'' which is a tool for
simulating continuous real-time models \cite{OM00}. 

Our approach is different. First, we do not intend to conceive a tool
for {\em simulating} real-time systems, but for {\em verifying}
real-time systems. In other words, we do not intend to prototype
real-time systems but to prototype verification algorithms
for real-time systems.

Second we focus on {\em Timed Automata}. Since verification of hybrid
systems is undecidable in the general case \cite{ACH+95}, any
verification tool must restrict to some decidable class of real-time
systems, or must be authorised to diverge for some systems.  Timed
automata is a class of continuous real-time systems which is known to
be decidable \cite{AD94}. Real-Time Maude falls in
the second approach in the sense that the ``find'' strategy
implemented in this tool gives only {\em partially} correct answers
\cite{OM00}.

The implemented model-checking algorithms for timed automata are
typical examples where the combination of exploration rules, deduction
rules, constraint solving and decision procedures are needed.  One aim
of this work is to argue and demonstrate through this example that the
rewriting calculus is a natural and powerful framework to understand
and formalise combinations of proving and constraint solving
techniques. 

Another aim is to argue the suitability advantages of using a formal
tool such as \ELAN\ to specify and prototype a model checking algorithm
compared to doing it in a much cumbersome way using a conventional
programming language. First this allows a clear flexibility for
customisation not available in typical hard-wired model checkers,
through for example programmable strategies.  Second, using the
efficient \ELAN\ compiler, the performances are indeed close to
dedicated optimised model-checking tools.

This paper is organised as follows. In Section \ref{elan}, we describe
the \ELAN\ system based on rewriting calculus. In Section
\ref{timedautomata}, we recall what a timed automaton is. In Section
\ref{tool}, we describe our tool for verifying reachability
properties of product of timed automata. In Section
\ref{implementation}, we discuss the implementation.

\section{The \ELAN\ system}
\label{elan}

The \ELAN\ system takes from functional programming the concept of
abstract data types and the function evaluation principle based on
rewriting. In \ELAN, a program is a set of labelled conditional
rewrite rules with local affectations 
$$\ell: l \Rightarrow r \mbox{ if } c \mbox{ where } w$$
Informally, rewriting a
ground term $t$ consists of selecting a rule whose left-hand side
(also called pattern) matches the current term ($t$), or a subterm
($t_{|\omega}$), computing a substitution $\sigma$ that gives the
instantiation of rule variable ($l\sigma = t_{|\omega}$), and if instantiated
condition $c$ is satisfied ($c\sigma$ reduces to $true$), applying
substitution $\sigma$ enriched by local affectation $w$ to the right-hand
side to build the reduced term.

In general, the normalisation of a term may not terminate, or
terminate with different results corresponding to different selected
rules, selected sub-terms or non-unicity of the substitution $\sigma$. So
evaluation by rewriting is essentially non-deterministic and
backtracking may be needed to generate all results.

One of the main originalities of the \ELAN\ language is to provide
strategies as first class objects of the language. This allows the
programmer to specify in a precise and natural way the control on the
rule applications. This is in contrast to many existing
rewriting-based languages where the term reduction strategy is
hard-wired and not accessible to the designer of an application. The
strategy language offers primitives for sequential composition,
iteration, deterministic and non-deterministic choices of elementary
strategies that are labelled rules. From these primitives, more
complex strategies can be expressed, and new strategy operators can be
introduced and defined by rewrite rules.

The full \ELAN\ system includes a preprocessor, an interpreter, a
compiler, and standard libraries available through the \ELAN\ web
page\footnote{{\tt http://elan.loria.fr}}. From the specific
techniques developed for compiling strategy controlled rewrite
systems~\cite{Moreau-RTA00,MoreauK-PLILP+ALP98}, the \ELAN\ compiler
is able to generate code that applies up to 15 millions rewrite rules
per second on typical examples where no non-determinism is involved
and typically between 100 000 and one million controlled rewrite per
second in presence of associative-commutative operators and
non-determinism.

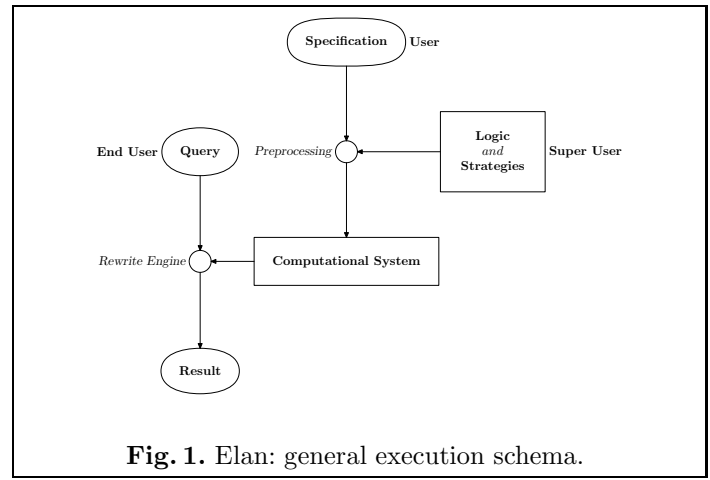
\begin{figure}[hbt]
\begin{center}
\framebox{\parbox{9cm}{
$$\epsfig{figure=schema.1, width=7cm}$$
\caption{Elan: general execution schema. \label{elan-principle}}
}}
\end{center}
\end{figure}

\section{Timed automata}
\label{timedautomata}

A \motnouv{clock} is a variable which takes value in the set $\R^+$ 
of non-negative real numbers.  A {\it clock constraint} is a
conjunction of constraints of type $x \# c$ or $x-y \# c$ for some
clocks $x,y$, rational number $c \in \Q$ and $\#
\in \{\leq,<,=,>,\geq\}$. Let $TC(K)$ denotes the set of clock constraints over
clock set $K$.

Informally, a \motnouv{Timed Automaton} is a finite automaton
augmented with clock variables, which can be reset to $0$ by guarded
transitions.

\begin{figure}[htbp]
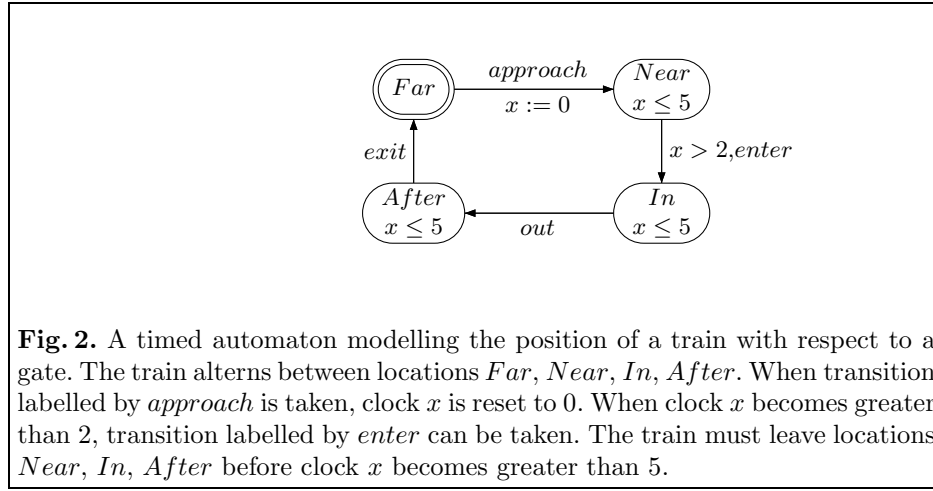

\noindent\framebox{\parbox{\textwidth}{
\begin{automaton}{50}{50}{h}{3cm}
\initialnode(s0)(0,50){\conc{$Far$}{}}
\node(s1)(100,50){\conc{$Near$}{$x\leq 5$}}
\node(s3)(0,0){\conc{$After$}{$x\leq 5$}}
\node(s2)(100,0){\conc{$In$}{$x\leq 5$}}
\edgecond{s0,s1}{$approach$}{$x:=0$}
\edgecond{s1,s2}{$x>2$,$enter$}{}
\edgecond{s2,s3}{$out$}{}
\edgecond{s3,s0}{$exit$}{}
\end{automaton}
\caption{A timed automaton modelling the position of a train with
  respect to a gate. The train alterns between locations $Far$,
  $Near$, $In$, $After$. When transition labelled by $approach$ is
  taken, clock $x$ is reset to $0$.  When clock $x$ becomes greater
  than $2$, transition labelled by $enter$ can be taken. The train must
  leave locations $Near$, $In$, $After$ before clock $x$ becomes greater than
  $5$.}
 \label{figtrain}
}}
\end{figure}

Formally, a \motnouv{timed automaton} \cite{AD94,Alu98}
is a $5$-tuple
$\mathcal{A}=(\Sigma,L,K,I,\Delta)$ where
\begin{enumerate}
\item $\Sigma$ is a finite alphabet,
\item $L$ is a finite set of \motnouv{locations},
\item $K$ is a finite set of clocks. A \motnouv{state} is given by some location and
some valuation of the clocks, i.e. by some element $(s,v)$ of $L \times
\R^{+^{|K|}}$.
\item $I$ is a function from $L$ to $TC(K)$ that labels each location
  $s$ by some \motnouv{invariant} $I(s)$. Invariant $I(s)$ restricts
  the possible values of the clocks in location $s$. 
\item $\Delta$ is a subset of $\Sigma \times L \times TC(K) \times \mathcal{P}(K) \times L$.
  \motnouv{Transition} $(a,s,c,z,s') \in \Delta$ corresponds to a
  transition from location $s$ to location $s'$, labelled by 
  $a$, guarded by constraint $c$ on the clocks, and which resets the
  clocks $k \in z$ to $0$.
\end{enumerate}

Timed automaton $\mathcal{A}$ evolves according to two basic types of transitions:
\begin{enumerate}
\item \motnouv{Delay transitions} corresponds to the elapsing of time
  while staying in some location: write $(s,v) \longrightarrow^{d} (s,v')$, where
  $d \in \R^+$, $v'=v+(d,\dots,d)$ provided for every $0 \leq e \leq d$,
  state $v+(e,\dots,e)$ satisfies constraint $I(s)$.
\item \motnouv{Action transitions} corresponds to the execution of
  some transition from $\Delta$: write $(s,v) \longrightarrow^{a} (s',v')$, for $a \in
  \Sigma$, provided there exists $(a,s,c,z,s') \in \Delta$ such that $v$
  satisfies constraint $c$ and $v'_{k}=v_{k}$ for $k \not\in z$,
  $v'_{k}=0$ for $k \in z$.
\end{enumerate}
A \motnouv{trajectory} of $\mathcal{A}$ starting from $(s,v)$ is a
sequence $$(s,v) \longrightarrow^{e_0} (s_1,v_1) \longrightarrow^{e_1} (s_2,v_2) \dots$$ for some
$e_0,e_1,\dots \in \Sigma \cup \R^+$.

\begin{figure}[htbp]
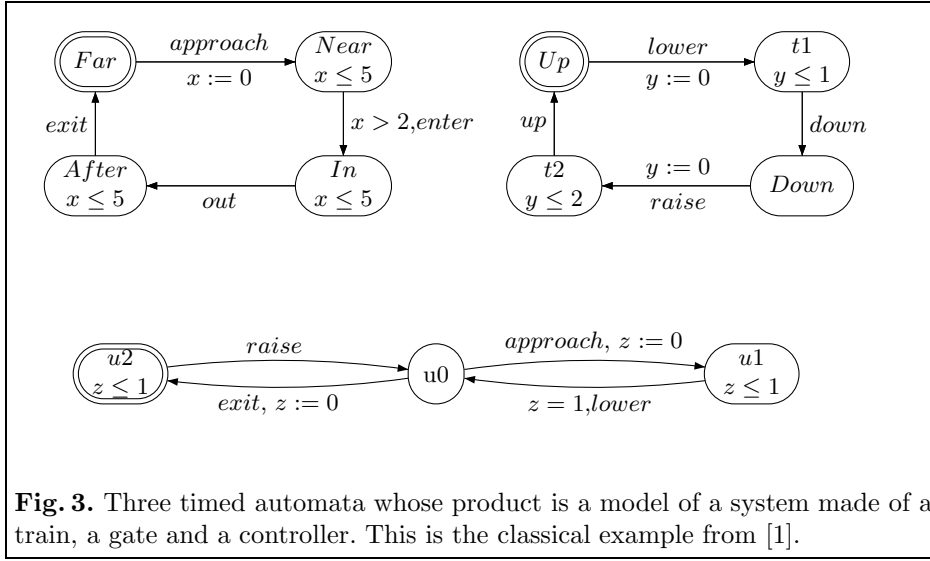

\noindent\framebox{\parbox{\textwidth}{
\hspace{-2cm}
\begin{tabular}{cc}
\begin{minipage}{6.5cm}
\begin{automaton}{20}{50}{h}{3cm}
\initialnode(s0)(0,50){\conc{$Far$}{}}
\node(s1)(100,50){\conc{$Near$}{$x\leq 5$}}
\node(s3)(0,0){\conc{$After$}{$x\leq 5$}}
\node(s2)(100,0){\conc{$In$}{$x\leq 5$}}
\edgecond{s0,s1}{$approach$}{$x:=0$}
\edgecond{s1,s2}{$x>2$,$enter$}{}
\edgecond{s2,s3}{$out$}{}
\edgecond{s3,s0}{$exit$}{}
\end{automaton}
\end{minipage}  & 
\begin{minipage}{5.5cm}
\begin{automaton}{20}{50}{h}{3cm}
\initialnode(t0)(0,50){\conc{$Up$}{}}
\node(t1)(100,50){\conc{$t1$}{$y\leq 1$}}
\node(t3)(0,0){\conc{$t2$}{$y \leq 2$}}
\node(t2)(100,0){\conc{$Down$}{}}
\edgecond{t0,t1}{$lower$}{$y:=0$}
\edgecond{t1,t2}{$down$}{}
\edgecond{t2,t3}{$raise$}{$y:=0$}
\edgecond{t3,t0}{$up$}{}
\end{automaton}
\end{minipage} \\
\multicolumn{2}{c}{
\begin{minipage}{8cm}
\begin{automaton}{100}{10}{h}{3cm}
\initialnode(u2)(0,0){\conc{$u2$}{$z\leq 1$}}
\node(u0)(70,0){u0}
\node(u1)(140,0){\conc{$u1$}{$z\leq 1$}}
\edgecondcurve{u0,u2}{$exit$,}{$z:=0$}
\edgecondcurve{u2,u0}{$raise$}{}
\edgecondcurve{u0,u1}{$approach$,}{$z:=0$}
\edgecondcurve{u1,u0}{$z=1$,$lower$}{}
\end{automaton}
\end{minipage}}
\end{tabular}
\caption{Three timed automata whose product is a model of a system made 
  of a train, a gate and a controller. This is the classical example
  from \cite{Alu98}.}  \label{figproduit} }}
\end{figure}

\paragraph{Product construction.} Timed automata can be composed by
\motnouv{synchronisation} \cite{Alu98,AD94} (see Figure \ref{figproduit}
for a classical example).  Intuitively, building the
product of two timed automata consists in considering a timed
automaton whose state space is the product of the state spaces of the
automata, where transitions labelled by a same letter in the two
automata occur synchronously, and others may occur asynchronously.





\section{The tool}
\label{tool}

\begin{figure}[hbt]
\framebox{\parbox{9cm}{
\begin{center}
$$\epsfig{figure=schema.3, width=7cm}$$
\caption{Execution schema of the tool.
  \label{fig:schem}}
\end{center}
}}
\end{figure}

Our verification system for timed automata is fully implemented in
\ELAN. It works according to the schema of Figure \ref{fig:schem}.
Concretely, 
\begin{enumerate}
\item The tool takes a specification of a product of
  automata. The specification is given in the form of
  a text file containing a list of clocks, a list of locations, a list
  of labels, and a list of automaton descriptions. Each automaton
  description is in turn a list of list of locations, labels,
  invariants, transitions: cf Figure \ref{fig:spec}.
  
\item The specification is then parsed using the \ELAN\ system, and
  compiled into an executable program. More precisely, the \ELAN\ 
  preprocessor, manipulates the encapsulated lists of the
  specification through \ELAN\ rules in order to generate rewrite
  rules, which are in turn compiled by the \ELAN\ compiler into an
  executable $C$ program.

\item This $C$ program tests reachability properties of the product of
automata: 
\begin{enumerate}
  \item it takes as input a query of type $go(s/c, s'/c')$ where
$s,s'$ are some locations of the product of
automata, and $c,c'$ are some clock
constraints. 
\item it answers $True$ iff there is a trajectory starting from some
  state $(s,v)$ with $v$ satisfying clock constraint $c$ that reaches
  some $(s',v')$ with $v'$ satisfying clock constraint $c'$.
\end{enumerate}
\end{enumerate}

\begin{figure}
{\scriptsize
\begin{tabular}{|c|c|}
\hline
\begin{minipage}{6.5cm}
\begin{verbatim}
specification train
  Clocks
    X Y Z
    nil
  States
    Far Near In After Up t1 Down t2 u0 u1 u2
    nil
  Labels
    app raise lower up down enter out exit
    nil
  Automata
    (
    Locations
      Far Near In After
      nil
    Labels
      app enter out exit
      nil
    Invariants
      Far   :        true
      Near  : X<=5 ^ true
      In    : X<=5 ^ true
      After : X<=5 ^ true
      nil
    Transitions
      Far  , app  :        true, X nil, Near  .
      Near , enter: X>2  ^ true,   nil, In    .
      In   , out  :        true,   nil, After .
      After, exit :        true,   nil, Far   .
      nil
    ) .
(
    Locations
      Up t1 Down t2
      nil
    Labels
\end{verbatim}
\end{minipage}
& 
\begin{minipage}{6.5cm}
\begin{verbatim}    

      lower down raise up
      nil
    Invariants
      Up   :        true
      t1   : Y<=1 ^ true
      Down :        true
      t2   : Y<=2 ^ true
      nil
    Transitions
      Up  , lower :        true, Y nil, t1   .
      t1  , down  :        true,   nil, Down .
      Down, raise :        true, Y nil, t2   .
      t2  , up    :        true,   nil, Up   .
      nil
    ) .
    (
    Locations
      u0 u1 u2
      nil
    Labels
      app lower raise exit
      nil
    Invariants
      u0 : true
      u1 : Z<=1 ^ true
      u2 : Z<=1 ^ true
      nil
    Transitions
      u0, app   :               true, Z nil, u1 .
      u1, lower : Z>=1 ^ Z<=1 ^ true,   nil, u0 .
      u0, exit  :               true, Z nil, u2 .
      u2, raise :               true,   nil, u0 .
      nil
    ) .
    nil
end
\end{verbatim}
\end{minipage} \\
\hline
\end{tabular}
}
\caption{The specification of the system of Figure
  \ref{figproduit}. Each automaton is described by lists of locations,
  labels, invariants and transitions. The product is in turn described
  by a list of automata.}
\label{fig:spec}
\end{figure}

Figure \ref{figexample} shows some queries and their execution times for
the $train$ $\|$ $gate$ $\|$ $controller$ system, and for the system
described in \cite{DY95} consisting of two robots, a conveyer belt and a
processing station\footnote{See web page {\tt http://www.loria.fr/$\sim$kacem/AT}
  for details.}.  The execution times of  system
\motnouv{KRONOS} are shown for comparison. Observe that they are of
same magnitude.

\begin{figure}
{
\hspace{-1cm}
\begin{tabular}{|l|l|c|c|}
\hline
Query & Answer &   {\sf Kronos} & {\sf Elan} \\
\hline \hline
$Train$: $go(Far.Up.u0.nil/true,In.Down.u0.nil/true)$ & $True$ 
 & 0.03 seconds & 0.00 seconds
\\
\hline

$Train$: $go(Far.Up.u0.nil/true,In.Up.u0.nil/true)$ & $False$  & 0.03
seconds & 0.03 seconds
\\
\hline
$Robots$:$go(D-Wait.G-Inspect.S-Empty.B-Mov.nil/true$
                                & $True$ & 0.04 seconds & 0.03 seconds
\\
$,D-Turn-L.G_-Inspect.S-Empty.B-Mov.nil/true)$ & & & \\ \hline

$Robots$:
$go(D-Wait.G-Inspect.S-Empty.B-Mov.nil/true,$
& $False$ & 0.04 seconds & 0.14 seconds  \\
$D-Turn-R.G-Turn-L.S-Empty.B-On-S.nil/true)$ & & & \\
\hline
\end{tabular}
}
\caption{Some queries and their  execution
  times
  (performed on a PC PIII 733 MHz 128 Mo).} \label{figexample}
\end{figure}

\section{Implementation}
\label{implementation}

We now describe how the reachability algorithm is implemented using
rewrite rules controlled by strategies. We need first to recall the
basis of timed automata theory.

\subsection{Region automaton}

Recall that a \motnouv{labelled transition system} is a tuple
$(Q,\Sigma,\to)$ where $Q$ is set of states, $\Sigma$ is some finite alphabet,
and $\to \in Q \times \Sigma \times Q$ is a set of transitions.

Given some timed automaton $A=(\Sigma,L,K,I,\Delta)$, denoting $(s,v) \leadsto^{a}
(s',v')$ if there exists $s''$ and $v''$ such that $(s,v) \longrightarrow^{d}
(s'',v'') \longrightarrow^{a} (s',v')$ for some $d \in \R^+$, the
\motnouv{time-abstract labelled transition system} associated to $A$
is the labelled transition system $S(A)=(Q_A,\Sigma,\leadsto)$ whose state space
$Q_A$ is unchanged, i.e. $Q_A=L \times \R^{+^{|K|}}$, but whose transition
relation is given by $\leadsto$.

Although $S(A)$ has uncountably many states,  we can associate some equivalence relation $\part_A$ over the
state space $Q_A$ which is stable and which is of finite index
\cite{Alu98,AD91}.  Some equivalence relation $\part$ over
the space of a labelled transition system $(Q,\Sigma,\to)$ is said to be
\motnouv{stable} iff whenever $q \part u$ and $q \to^{a} q'$ there
exists some $u'$ with $u \to ^{a} u'$ and $u' \part u$.

The \motnouv{quotient of $S(A)$ with respect to $\part_A$}, denoted by
$[S(A)]$, is the transition system whose state space is made of the
equivalence classes of $\part_A$, called \motnouv{regions}, and such
that there is a transition from region $\pi$ to region $\pi'$ labelled
by $a$ if for some some $q \in \pi$ and $q' \in \pi'$, $q \leadsto^{a} q'$.

Since $\part_A$ is of finite index, $[S(A)]$ is a finite automaton, which
is called \motnouv{the region automaton} of $A$ \cite{Alu98,AD91}.

Since $\part_A$ is stable, the set of reachable states from some
region $s_0$ in timed automaton $A$ is equal to the union of the
reachable regions in $[S(A)]$ starting from region $s_0$
\cite{Alu98,AD91}.  Hence, the reachability problem for $A$ reduces to
the reachability problem for finite automaton $[S(A)]$.

This is the basis of all model-checking tools for timed automata. See 
\cite{Alu98,AD94} for details.

\subsection{Manipulating regions using states with constraints}

Computing the reachable regions in region automaton $[S(A)]$ requires
to manipulate regions. 

This is can be done by manipulating symbolic representations of these
sets, i.e. by manipulating clock constraints \cite{Alu98,AD94}. Hence,
our program in \ELAN\ manipulates terms of form $s/c$ where $s$, $c$
is a (term representation of) a state of the product of automata, and
$c$ is a clock constraint.  Term $s/c$, represents the (convex) set, denoted by
$\lsemantics s/c\rsemantics$, of all the states $(s,v) \in Q_A$, such that $v$ satisfies
constraint $c$. Such a set is called a \motnouv{zone}
\cite{Alu98,AD94}.

The heart of the reachability algorithm in \ELAN\ is made of {\em
  rewrite} rules which manipulate such terms through symbolic
operators on constraints. Figure \ref{figconstraint} shows an example
of such a rule for the system of Figure \ref{fig:spec}. This rewrite
rule uses ``Intersection'' and ``Reset'' operators on clock
constraints.

\begin{figure} 
{\scriptsize
\begin{verbatim}
[] Post.enter(Near/c) => In /Reset(Intersection(X>2 ^ true,c), nil) end        
\end{verbatim}
}
\caption{A rewrite rule manipulating a zone, using
`Intersection'' and ``Reset'' operators on clock
constraints.} 
\label{figconstraint}
\end{figure}

The $Intersection$ operator transforms two clock constraints into a
representation of their intersection.  The $Reset$ operator transforms
a constraint $c$, and a list of clocks $k$, to a constraint
representing the set of states reachable from a state satisfying $c$
after the variables of $k$ are reset to $0$.

\subsection{Generation of rules from the automaton specification}

These rules on zones are generated from the description of the timed
automaton given as input using the preprocessor facilities of the
\ELAN\ system.


For example, for any transition $e=(a,s,d,z,s')$ of the timed
automaton, we need to generate a rewrite rule which transforms any
zone $s/c$ into some zone $s'/c'$ which represents all the states
reachable from $s/c$ by transition $e$. Formally, we want to rewrite  
$s/c$ into $s'/c'$ with 
$$\lsemantics s'/c'\rsemantics=\{(s',v')| (s,v) \in \lsemantics s/c\rsemantics \mbox{ and } (s,v)
\longrightarrow^{a}(s',v') \}$$
This can be done by generating rewrite rule
$$post.e(s/c) \Rightarrow s'/Reset(Intersection(c,d),z)$$

In order to generate this rewrite rule for any transition $e$ of the automaton
specification, we just need in the \ELAN\ system, to write the
preprocessor rule of Figure \ref{figpreproc}.

\begin{figure}
{\scriptsize
\begin{verbatim}
FOR EACH A : Automaton SUCH THAT A:=(listExtract) elem(LA):{
rules for statezone
  z : clockzone ;
 global
  FOR EACH tr       : transition ;
           bef, aft : state ;
           label    : label ;
           cond     : clockzone ;
           zero     : list[clock]
  SUCH THAT tr   := (listExtract) elem(lst_trans(A))
        AND bef  :=()tr_before(tr)
        AND label:=()tr_label (tr)
        AND cond :=()tr_cond  (tr)
        AND zero :=()tr_zero  (tr)
        AND aft  :=()tr_after (tr) :
  {
    []  Post.label(bef/z) => aft / Reset(Intersection(cond,z),zero) end
  }
end // of rules for statezone
\end{verbatim}
  }
\caption{The preprocessor rule which generated the rule of Figure
  \ref{figconstraint} for the transition from location $Near$ to $In$ 
  of the system of Figure \ref{figproduit}.} \label{figpreproc}
\end{figure}

\subsection{Constraint representation}

Implementing the operators on constraints requires to represent clock
constraints.

One solution consists in representing clock constraints using
\motnouv{bounded differences matrices} \cite{Alu98,Dil89}.  With this
representation scheme, any clock constraint has a normal form which
can be computed using a Floyd-Warshal algorithm based technique
\cite{Alu98}. 

This method using bounded differences matrices has been implemented as
a rewrite system in our tool. That means in particular that the
Floyd-Warshal based technique for computing normal forms of bounded
differences matrices is implemented as rewrite rules, as well as all
operators required on constraints ($Intersection$, $Reset$,
$Is\_Empty?$ , $Is\_Equivalent?$, $Effect\_Of\_Time$$\_Elapsing$).

\paragraph{Constraint representation alternatives.} 
Other representations of constraints are possible. In particular, we
have experimented the representation of clock constraints using
classical logical formulas.  Clock constraints are closed by
quantifier elimination \cite{BeffaraStage2000,LivreClarke}. Denoting
by $Exists$ the operator which maps a clock constraint $c$ and a list
of variable $k$ to a clock constraint logically equivalent to formula
$\exists k\ c$, the above operators on constraints can all be expressed
using the $Exists$ operator \cite{BeffaraStage2000,LivreClarke}. For
example, the $Reset$ operator can be expressed by the rewrite rule
$$Reset(c,k) \Rightarrow Exists(c,k) \land k=0.$$

This as been implemented in our tool. The $Exists$ operator is
computed using a technique based on Fourier-Motzkin algorithm
\cite{BeffaraStage2000}.

\subsection{Exploration}

Once the constraint manipulation is implemented, one interesting part
is to implement the exploration of the reachable regions of the
automaton. This is done by manipulating terms of the form
$Transitions\_List(lsz,szs,szc)$ where $lsz$ is a list of already
explored zones, $szs$ is the current zone under investigation, and
$zsc$ is the objective zone.

One main originality of \ELAN\ is the possibility to
express strategies.  Hence, the exploration of graph can be guided by
the simple strategy language of the \ELAN\ system.

As an example, suppose we want to explore the graph by backtracking.
Rule named $SuccessStep$ of Figure \ref{figstep} does one step of the
graph exploration for the particular case when the objective zone is
reachable by one step of the graph exploration, and rule named
$NextStep$ does one step of the graph exploration for the generic
case.

To explore the whole graph, we just need to iterate these rules,
taking at each step the first one of the two elementary rules
$SuccessStep$ and $NextStep$ which succeeds. This is easily done using 
the \ELAN\ strategy language as in Figure \ref{fig:strategy}.

\begin{figure}
{\scriptsize
\begin{verbatim}



rules for bool
  s,sO : state;
  sz      : zone;
  c,cO : clockzone;
  result  : bool;
  lsz : hashSet[statezone];
global
[SuccessStep] Transitions_List(lsz,s/c,s0/c0) => result
                  where sz := (Post) s/c
                  if sz.state== s0
                  if not Is_Empty?(Intersection(sz.constraint,c0))
                  where result:=() True 
                  end
[NextStep]    Transitions_List(lsz,s/c,s0/c0) => result
                  where sz := (Post) s/c
                  if not Is_Empty?(sz.constraint) 
                  if not lsz.contains(sz) 
                  where result:=(Exploration) Transitions_List(lsz.add(sz),sz,s0/c0)
end
\end{verbatim}
}
\caption{Making one step of the exploration.} \label{figstep}
\end{figure}

\begin{figure}
{\scriptsize
\begin{verbatim}



strategies for bool
 implicit
  []  Exploration => first one(SuccessStep,NextStep) end
  end


rules for bool
  s,sO: state;
  c,cO: clockzone;
  result: bool;
global
[] go(s/c,s0/c0) => result 
               choose
                 try where result:=(Exploration) Transitions_List(EmptySet,s/c,s0/c0)
                 try where result:=()False
               end
end


\end{verbatim}
}
\caption{Exploring the whole graph.} \label{fig:strategy}
\end{figure}

\paragraph{Exploration alternatives.} Of course, other exploration
strategies can also be used and experimented just by modifying the
above lines. Graph can easily be traversed depth first, breath
first, if one prefers.

\subsection{On-fly generation.}  

The tool implements \motnouv{on-fly model-checking}. This means that
the tool does not need to build the full product of the timed automata
before testing reachability properties, but that the transitions of
the product of timed automata are generated on-line only when needed.
This is in contrast with what happens in some model-checking tools.

This is easily done, in the case of an input consisting of a product
of $n$ timed automata, by using a succession of $n$ \ELAN\ $first\_one$
strategy operators applied on named rules $ExecuteTransition\_i$ and
$JumpStep\_i$ which compute on-line the transitions to apply in each
automaton of the product corresponding to some possible label: cf Figure
\ref{figonline}.

\begin{figure}
{\scriptsize
\begin{verbatim}
// Transcription Of The Synchronised Product
FOR EACH N : int SUCH THAT N:=()size_of_Automaton_list(LA):{
rules for statezone
  {s_I : state ;}_I=1...N
  {ss_J : state ;}_J=1...N
  Phi, nPhi       : clockzone ;
  lbl             : label ;
  sz              : statezone ;
 global
{
  [ExecuteTransition_i] SynTransitionOperator(lbl,{s_j.}_j=1...(i-1) s_i.{s_j.}_j=(i+1)...N nil/Phi) =>
         SynTransitionOperator(lbl,{s_j.}_j=1...(i-1) ss_i.{s_j.}_j=(i+1)...N nil/nPhi)
                                      if action(lbl,s_i,i)
                                      where sz:=()TransitionOperator.lbl(s_i/Phi)
                                      where ss_i:=()st(sz)
                                      where nPhi:=()zn(sz)
                                      end
  [JumpStep_i] SynTransitionOperator(lbl,sz) => SynTransitionOperator(lbl,sz) end
}_i=1...N
  [FinishSynTransition]   SynTransitionOperator(lbl,sz) => sz end
end // of rules for statezone
} // End Of The Transcription Of The Synchronised Product

strategies for statezone
 implicit
 FOR EACH N : int SUCH THAT N:=()size_of_Automaton_list(LA) :{
  [] next_sz => {first one(ExecuteTransition_I,JumpStep_I);}_I=1...N  FinishSynTransition end
 }
end // of strategies for statezone
\end{verbatim}}
\caption{On-line generation of the transitions of the product of
  automata. \label{figonline}} 
\end{figure}

\section{Conclusion}

This paper presents the use of the rewrite based system \ELAN\ to
prototype model-checking algorithms for timed automata. As expected,
the performance are a little bit lower than model-checking dedicated
tools like KRONOS \cite{DOTY95} or UPPAAL \cite{LPY97}. However, using
the specific techniques already developed for compiling strategy
controlled rewrite systems implemented in the \ELAN\ system compiler,
the performances turns out to be of same magnitude.

The main advantage is the gained flexibility compared to conventional
programming languages. The whole \ELAN\ code for the described
model-checkers is less than $1100$ lines (including comments).
Changing the graph exploration technique, or the constraint solving
algorithm, for example, turned out to require to modify only a few
lines. We presented the direct and classical implementation of the
reachability algorithms for timed automata. But other techniques
(e.g.: partition refinement techniques \cite{Alu98,ACH+95},
alternative representations for constraints
\cite{ABK$^+$97,BLP$^+$99,BM00,MLAH99b,W00}) could also be
experimented and would require only a few modifications of the
existing \ELAN\ code.

Furthemore, the \ELAN\ system offers facilities, such as a strategy
language which provides flexibily for customizations that are not
available in typical hard-wired model checkers such as programmable
strategies.

Moreover, we believe that such a work helps to understand mixture of
proving and constraint manipulation techniques by studying them in the
same unifying framework. In particular, it clearly helps to delimit
the difference between pure computations and deductions in
model-checking techniques which are very often presented as relying
only on computations on constraints. 

More details on the tool together with the full code can be found on
web page {\tt http://www.loria.fr/$\sim$kacem/AT}. 

\section{Thanks}

The authors would like to thank all members of the PROTHEO group for
their discussions.  Among them, we would like to address some special
thanks to Pierre-Etienne Moreau who helped us a lot through his
expertise of the \ELAN\ compiler.

\end{document}